# *Short term solar energy prediction by machine learning algorithms*

Shahid F.[1], Zameer A.[2*], Afzal M.[3], Hassan M.[4]
[1,2,3,4]Department of Computer & Information Sciences,
[1]E-mail: FarahShahid_17@pieas.edu.pk;
[2]E-mail: aneelaz@pieas.edu.pk;
[3]E-mail: mudasserch1@gmail.com;
[4]E-mail: muhammad.hassan.99.cs@gmail.com
Pakistan Institute of Engineering & Applied Sciences (PIEAS), Nilore, Islamabad 45650, Pakistan

*Abstract*— Smooth power generation from solar stations demand accurate, reliable and efficient forecast of solar energy for optimal integration to cater market demand; however, the implicit instability of solar energy production may cause serious problems for the smooth power generation. We report daily prediction of solar energy by exploiting the strength of machine learning techniques to capture and analyze complicated behavior of enormous features effectively. For this purpose, dataset comprising of 98 solar stations has been taken from energy competition of American Meteorological Society (AMS) for predicting daily solar energy. Forecast models of base line regressors including linear, ridge, lasso, decision tree, random forest and artificial neural networks have been implemented on the AMS solar dataset. Grid size is converted into two sections: 16×9 and 10×4 to ascertain attributes contributing more towards the generated power from densely located stations on global ensemble forecast system (GEFS). To evaluate the models, statistical measures of prediction error in terms of RMSE, MAE and R2_score have been analyzed and compared with the existing techniques. It has been observed that improved accuracy is achieved through random forest and ridge regressor for both grid sizes in contrast to all other proposed methods. Stability and reliability of the proposed schemes are evaluated on a single solar station as well as on multiple independent runs.

## Highlights

- Multivariate Analysis of Daily Solar Power Prediction in carried out.
- AMS dataset is used for two grid sizes.
- Linear, ridge, lasso, random forest, and artificial neural networks are used as regressors.
- Proposed ML schemes are evaluated on the basis of standard statistical error measures.
- Variants of ML techniques on two grid sizes of solar stations are employed.
- Comparison of error measures from proposed techniques is carried out with the existing techniques.
- Proposed techniques are also employed on a single solar station for robustness and evaluation.
- Random forest and ridge regressor furnish enhanced solar power prediction.

---

[*]Corresponding Author: Zameer A.

| Abbreviation | | | |
|---|---|---|---|
| RMSE | root mean square error | MLP | Multilayer perceptron |
| ANN | artificial neural network | NN | neural network |
| ARIMA | Autoregressive integrated moving average | LSR | least square regression |
| WGET-SWH | water-in-glass evacuated tube solar water heater | CHA-CNN | chaotic hybrid algorithm-convolutional neural networks |
| AMS | American Meteorological Society | HTS | high-throughput screening |
| CNN | Convolutional neural network | AR | Autoregressive |
| DBM | Deep Boltzmann machine | SVR | support vector regression |
| ELM | Extreme learning machine | WPP | wind power prediction |
| EGB | Extreme gradient boosting | FFNN | feed forward neural network |
| ML | Machine learning | RBFNN | Radial basis function neural network |
| GPR | Gaussian process regression | GBR | Gradient Boosted Regression |
| LSTM | long short-term memory | MAE | Mean absolute error |
| RBF | radial basis function | LSSVM | least squares support vector machine |

# 1  Introduction

Renewable and sustainable energy resources like solar energy, wind energy and tidal energy are gaining attention due to global, economic and political dilemma along with alarming pollution level in the air, water and soil [1]. Sustainable energy is enough and favorable to environment; however, due to inherent fluctuations of sustainable energy, it seems very difficult to be integrated into power grids in the form of geographical and demographic position. Among different sustainable energies, solar energy turns into focal point in the field of energy industry having the potential to reduce carbon footprint and combat rising electricity costs [2]. The major problem with solar energy includes continuously changing and unpredictable weather, cloudy density, climate and seasons, which restricted the appliance of solar irradiation. This causes fluctuations in solar energy production. Thus, resource planner and companies are in search of models that accommodate these uncertainties for design and management of solar energy production on daily basis, which could allow them to fulfil demand and supply of consumers regardless of weather conditions [3]. Therefore, short-term solar energy prediction is intensely critical [4, 5].

These methods are divided into sub groups of physical, statistical methods and mix of these models [6, 7]. Physical methods employ various physical variables of location obstacles and atmospheric behavior to gather local wind speed for wind power estimation [8], [9], [10]. McPherson *et al.* considered the observations to intensify the knowledge of physical processes that arise covering spatial and temporal scales [11]. Statistical energy prediction models consist of set of expressions and help in the construction of computational models to obtain desired results [12, 13]. With the fast development, machine learning techniques grabbed the attention in several fields of acoustic disaster [14], air-water based thermoelectric cooling unit [15], sense-based text [16], speech emotion recognition [17], electricity price forecasting [18], functional magnetic resonance imaging [19], photovoltaic power prediction [20], wind power prediction [21] and cancer prediction [22] [23]. These models generated by machine learning algorithms learn from previous data to predict for future unseen data. Appropriate adaption of suitable machine learning models with respect to structure of renewable solar plants is crucial. Lauret *et al.* has discussed linear and non-linear statistical techniques autoregressive (AR) and artificial neural network (ANN), while employing support vector machine (SVM), respectively [24]. Liu *et al.* proposed a high-throughput screening (HTS) model based ANN to construct and screen water-in-glass evacuated tube solar water heater (WGET-SWH) [25]. Conventional machine learning methods of time series have been utilized to predict solar energy. Currently, SVM has been mostly implemented in several problems of renewable and sustainable energy power systems [26]. There is a number of studies in literature related to machine learning techniques for solar energy prediction, including Naïve Bayes forecasting method to predict probability of one hour ahead solar power [27]. Zeng *et al.* ascertained that SVM has good generalization ability with linear, polynomial and radial basis function based kernel; which demonstrate enhanced performance of SVM over AR and radial basis function neural network (RBFNN) [28]. Non-linear models, such as neural networks employ nonlinear mapping between input and output variables of solar energy as reported in [29]. A non-parametric method, Gaussian process regression (GPR) encodes features of solar energy and adopts flexible kernel function to provide confidence interval for prediction [30].

To find efficient and reliable solar energy forecast, many researchers are motivated to model optimally by incorporating the inherent fluctuations and complexity of solar energy. However, machine learning methods are taking roots for precise prediction [31]. Zameer *et al.* reported a hybrid technique based on ANNs and genetic programming to predict accurate short term wind power [32]. Chang *et al.* implemented hybrid approach of random forest regressor and Bayesian model to predict the solar energy [33]. Dong *et al.* has predicted solar energy through gradient boosted regression tree, convolutional neural networks (CNN), artificial neural networks (ANNs), K-means RBF and chaotic hybrid algorithm CHA-CNN [34]. Aggarwal *et* al. have reported feed forward neural network (FFNN) and regularized least square regression (LSR) on solar dataset, where FFNN produced better results [35]. Torres-Barran *et* al. have exploited Support Vector Regressor (SVR), Random Forest Regressor (RFR), Gradient Boosted Regressor (GBR) and Extreme Gradient Boosting (XGB) [36]. Ren et al. applied ensemble-forecasting methods of AdaBoost-ANN and EMD-AdaBoost-ANN for solar prediction [37]. Diaz-Vico *et* al. applied deep neural networks with RELU, dropout and weight decay for solar energy prediction [38].

We report exploration of solar energy characteristics and comparison of linear and non-linear machine learning methodologies to improve their generlization ability for better adaption and reduced prediction error. For this purpose, various machine learning based regression models are implemented for daily solar energy forecast. The key contributions of the presented work include:

- o Distribution of solar energy features on local and global scale for appropriate selection of grid size.
- o Implementation of ridge regression to reduce model complexity and multi collinearity.
- o Regularization of ridge regression to prevent overfitting by penalizing large weights.
- o Exploitation of lasso regression model with better feature selection capability.
- o Implementation of random forest regressor to capture the non-linear interactions between features and the target.
- o Employing artificial neural network with advantages over base line regressors in terms of learning from data samples rather than entire dataset to reduce the computation time.

Rest of the paper is organized as follows: in section II, the dataset and machine learning based proposed methodologies for solar energy prediction are described; in section III, experimental results and their comparison with existing techniques are discussed; finally, section IV concludes the overall work.

## 2 Materials and Methods

In this section, description of solar dataset and proposed machine learning based methodologies are presented. For evaluation of the proposed schemes, statistical performance measures are also defined.

### 2.1 Solar Energy Dataset

Dataset has been taken from a short-term solar energy prediction competition of American Meteorological Society (AMS) consisting of 98 solar stations to predict total daily solar energy. Oklahoma Mesonet site provided solar energy data on daily basis [39]. Table 1 dataset contains daily readings of a solar station from 01 January 1994 to 31 December 2007 of 98 solar stations. The dataset consists of the longitude and latitude distance and elevation distance in meters, for each station. The data gives solar energy in $MJ/m^{-2}$) and ha 15 attributes across each of 144 global ensemble forecast system (GEFS) locations on a 16 × 9 grid, which is 5113 days long. Moreover, the five sampling times are UTC (12:00, 15:00, 18:00, 21:00 and 24:00). The total forecast contains five-dimensional data of shape (5113, 11, 5, 9, 16). The 1$^{st}$ dimension shows date of the model on which it runs and gives some readings, while 2$^{nd}$ dimension gives the number of ensemble member for which the prediction is made. In this dataset, there are 11-ensemble members with perturbed initial conditions.

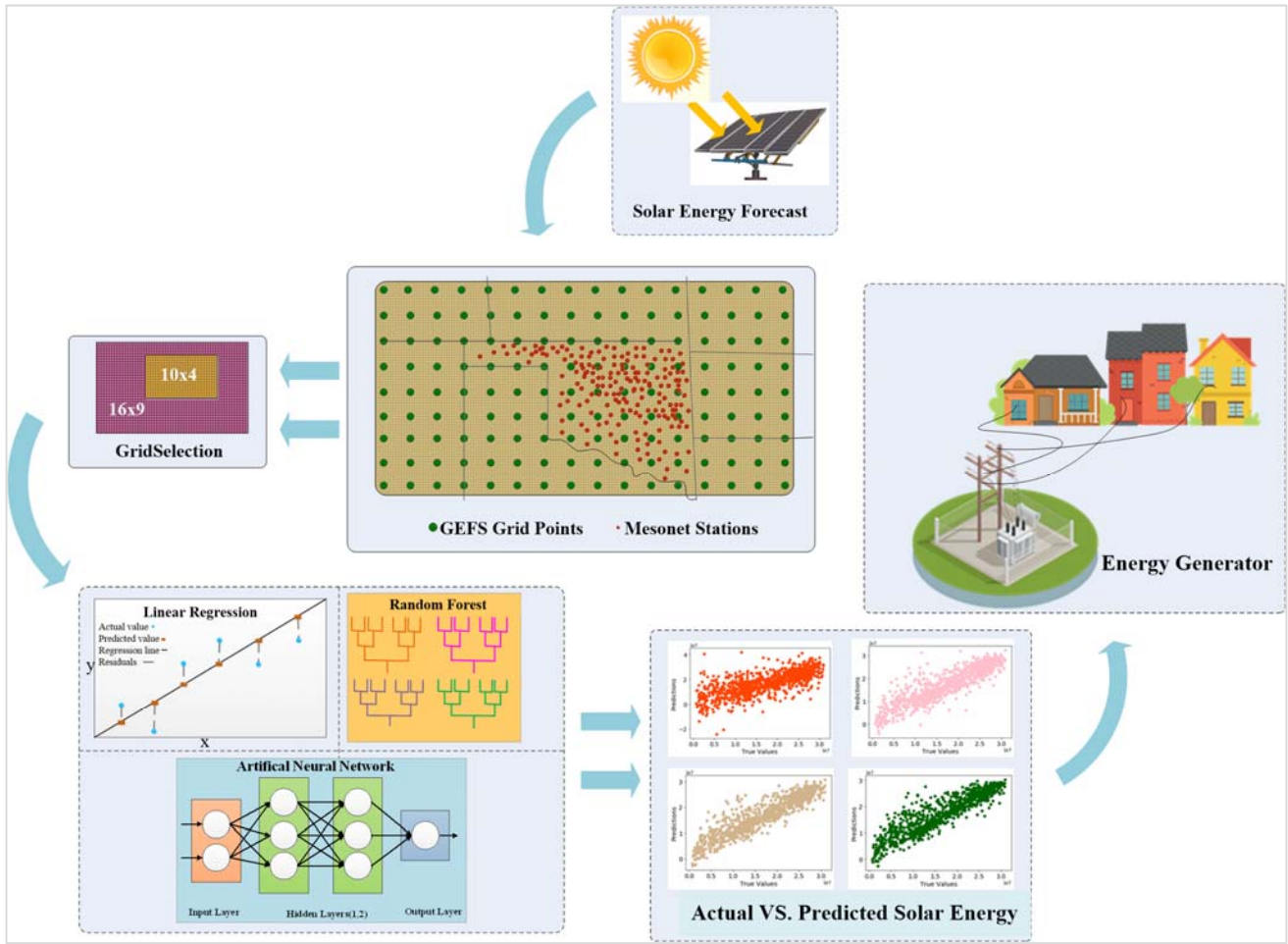

**Figure 1:** Graphical abstract representing the proposed models for solar energy prediction

**Table 1:** Detailed description of solar energy variables ref. [35]

| Variable | Description | Units |
|---|---|---|
| apcp_sfc | 3-Hour accumulated precipitation at the surface | kg/m$^2$ |
| dlwrf_sfc | Downward long-wave radiative flux average at the surface | W/m$^2$ |
| dswrf_sfc | Downward short-wave radiative flux average at the surface | W/m$^2$ |
| pres_msl | Air pressure at mean sea level | Pa |
| pwat_eatm | Perceptible Water over the entire depth of the atmosphere | kg/m$^2$ |
| spfh_2m | Specific Humidity at 2 m above ground | Kg |
| tcdc_eatm | Total cloud cover over the entire depth of the atmosphere | Kg$^{-1}$ |
| tcolc_eatm | Total column-integrated condensate over the entire atmos. | kg/m$^2$ |
| tmax_2m | Maximum Temperature over the past 3 hours at 2 m above the ground | K |
| tmin_2m | Minimum Temperature over the past 3 hours at 2 m above the ground | K |
| tmp_2m | Current temperature at 2 m above the ground | K |
| tmp_sfc | Temperature of the surface | K |
| ulwrf_sfc | Upward long-wave radiation at the surface | W/m$^2$ |
| ulwrf_tatm | Upward long-wave radiation at the top of the atmosphere | W/m$^2$ |
| uswrf_sfc | Upward short-wave radiation at the surface | W/m$^2$ |

The 3rd dimension represents the prediction hour, which ranges from 12-24 hours with an increment of 3 hours. These values begin from 00 UTC to depict same universal time to avoid any ambiguity. Whereby 4th and 5th dimensions denote the geographic coordinates of latitude and longitude in a uniform spatial grid. The comparison of GEFS grid points are shown in Figure 1 as green and the Mesonet stations as red.

**Table 2**: Parameter Settings of proposed ML methodologies

| Method | Parameters | Values |
| --- | --- | --- |
| Linear regression-I | Normalize | True |
| Linear regression-II | Normalize | False |
| Ridge regression-I | Alpha | 0.1 |
| Ridge regression-II | Alpha | 0.3 |
| Ridge regression-III | Alpha | 0.8 |
| Lasso regression-1 | Alpha | 0.1 |
| Lasso regression-II | Alpha | 0.3 |
| Lasso regression-III | Alpha | 0.8 |
| Decision tree-I | Max-depth | 2 |
| Decision tree-II | Max-depth | 7 |
| Random forest-I | Max-depth | 6 |
|  | Estimator | 20 |
| Random forest –II | Max-depth | 10 |
|  | Estimator | 40 |
| Random forest –III | Max-depth | 20 |
|  | Estimator | 60 |
| Artificial neural network-I | Optimizer | Adam |
| Artificial neural network –II | Optimizer | Rmsprop |

## 2.2 Grid and Feature Selection

According to the choice of features, we have reduced the grid size when selecting the GEFS points for better results. It is clear from the graphical diagram that most of the Mesonet stations have almost no solar power generation and whole grid of 16 × 9 that has not variety of values. Therefore, we have selected 10 × 4 grid size, where most of the solar stations reside for appropriate feature selection and to reduce the computational cost. Grid sizes 16 × 9 and 10 × 4 are taken as solar dataset A and solar dataset B, respectively.

## 2.3 Proposed machine learning approaches for short-term solar energy forecast

Machine learning techniques including Linear Regression, Ridge Regression, Lasso, Decision Tree, RFR are proposed for implementation on AMS datasets A & B. While neural network based regressor includes a simple sequential neural network. All these techniques are explained in the following subsections.

### 2.3.1 Multivariate Linear Regression

The technique of multivariate regression extends from the statistical model of linear regression in which multiple (two or more) independent variables $x$ is regressed for one dependent variable $y$ [40] [41]. In our case, there are $x_i = \{x_1, x_2, x_3, ..., x_n\}$ independent variables of solar energy dataset; where n is the number of inputs. Then, the estimated regression function is:

$$f(x_1, x_2, x_3, ..., x_n) = \beta_0 + \beta_1 x_1 + ... + \beta_n x_n + \varepsilon \tag{1}$$

Where the regression coefficients are $\{\beta_0, \beta_1, \beta_2, ..., \beta_n\}$ and random error is $\varepsilon$. The goal of regression function is to capture the dependencies between independent and dependent variables and to determine the values of predicted weights. For each sample, $i = \{1, 2, 3, ..., n\}$ of predicted response, $f(x_i)$ is close to actual response $y_i$. The difference among the predicted and actual response is called residuals, which is denoted as $y_i - f(x_i)$. Best predicted weights correspond to smallest residuals [42]. The ordinary least squares method is used to minimize the sum of squared residuals (SSR) to get the best weights. SSR is formulated as $SSR = \sum_i (y_i - f(x_i))^2$ for all samples.

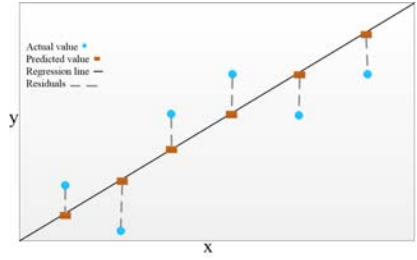

**Figure 2:** Regression line of Linear Regressor model

### 2.3.2 Ridge Regression

Ridge regression is the variant of linear regression [43, 44] and also known as regularization in which coefficients of models can be changed by changing the values of λ values. To reduce the bias of ridge regression estimators, Lawless *et* al. and Farebrother *et* al. introduced general biased estimators of regression coefficients [45, 46]. Eq (1) can be written in matrix form as:

$$\arg\min_{\beta \in R} \sum [y_i - f(x_i)] = \qquad (2)$$
$$\arg\min_{\beta \in R} \sum [y_i - (\beta_0 + \beta_1 x_1 + ... + \beta_n x_n)]$$

argmin attains the function to its minimum and find $\beta$'s that minimize the SSR. To make the ridge regression:

$$R^{ridge} = \arg\min_{\beta \in R} \sum_{i=1}^{n}(y_i - f(x_i))^2 + \lambda \sum_{j=1}^{p}(\beta_j)^2 \qquad (3)$$

From Eq (3), ridge regression is formed as:

$$R^{ridge} = \arg\min_{\beta \in R} \|y - x\beta\|^2 + \lambda \|\beta\|^2 \qquad (4)$$
$$\text{For} \quad \|\beta\|^2 < C$$

Here, subscription of $\|\beta\|^2$ is known as L2norm, $\|y - x\beta\|^2$ is the loss part and $\lambda\|\beta\|$ is the penalty part in equation (4). By iterating certain values of λ and evaluate the model with measurement of mean squared error (MSE). So, the value of λ that minimizes the error should selected as final model. The bigger value of λ cannot affect as $R^{ridge}$ can never be zero by the bigger value of λ.

$$\begin{cases} \lambda = 0 & \text{estimates the linear regression} \\ \lambda \geq 0 & \text{is tunning parameters} \\ \lambda = \infty & R^{ridge} = 0 \end{cases} \qquad (5)$$

### 2.3.3 Lasso Regression

Another machine learning technique implemented in this work is known as lasso, for "least absolute shrinkage and selection operator". It shrinks the magnitude of some regression coefficients and sets other coefficient to zero which helps in selection of features, which one to be accepted and/or ignored. The formula of lasso regression is:

$$L^{Lasso} = \arg\min_{\beta \in R} \sum_{i=1}^{n} \|y_i - f(x_i)\|^2 + \lambda \sum_{j=1}^{p} \|\beta_j\| \quad (6)$$

$$\text{where} \quad \|\beta\| \leq C$$

Equation (6) is used to minimize the sum of square residuals subject to L1 constraints, which causes many zeros that helps to pick up those features dependent on y. Here, C is the real number, which uses the L1 regularization so that value of regression coefficient is no more than C. We want to minimize the error by finding the best value of $\beta$.

### 2.3.3.1 Regularization

To reduce mean square error, some bias is added and very small values are chosen for the cost function to evaluate all models and their relative comparison. Another definition of regularization is that keeping the small number of parameters will not lead to too much error.

$$L_n \text{regularization} = \sum_i \left(|\theta_i^n|\right)^{1/p} \quad (7)$$

Lasso regression provides sparsity, L1 norm of lasso regression coefficient is in the form $\|\beta\|_1 = |\beta_0| + |\beta_1| + |\beta_2| + ... + |\beta_n|$, which shrinks the magnitude of regression coefficient exactly zero. L2 norm of ridge regression is written in the form $\|\beta\|_2 = \left(\beta_0^2 + \beta_1^2 + \beta_2^2 + ... + \beta_n^2\right)^{1/2}$, which shrinks the magnitude of regression coefficient near zero.

### 2.3.4 Decision Tree

The utmost general techniques in machine learning exploited for classification and regression problems is Decision trees [47], which is generated on If-Then rules to easily interpreted prediction that yields understandable models [48]. A decision tree is regression tree when numeric values are predicting, in which the top node is called the root node and all non-leaf node represents test on variables, each branch characterizes a result of the test, and all leaf node signifies a forecast [49].

### 2.3.5 Random Forest Regressor

Random forest produces many regression trees and each tree is made of a different bootstrap example from the original data using a tree regression algorithm [50]. After the forest is completed, a new object that needs to fit into regression line is put down each of the tree in the forest for regression. Each tree gives the numeric value that indicates the tree's decision about the object. The forest fits the object into regression line with the average of the individual tree predictions [51].

### 2.3.6 Artificial Neural Network

The conception of neural network largely imported from the topic of biology where neural network attempts capabilities of nervous systems to model the information. Neural Network is just composed of millions of neurons, which are each of them deals with incoming sample in many different ways [52], [53]. Each neuron takes a group of weighted inputs and reacts an output. The mathematical expression weighted inputs is given as.

$$a = \left(\sum_{i=1}^{n} w_i x_i\right) + b$$

Where, n is the number of input samples and $w_i$ are the weights of input and $b$ represents the bias of neuron. An activation function represented by $f$ is used process the weighted input and bias, compute output in the form is as below

$$f(a) = f\left[\left(\sum_{i=1}^{n} w_i x_i\right) + b\right]$$

Essentially, ANN model signifies those neurons having significantly excited inputs of big value. There are various type activation function like sigmoid and hyperbolic tangent function.

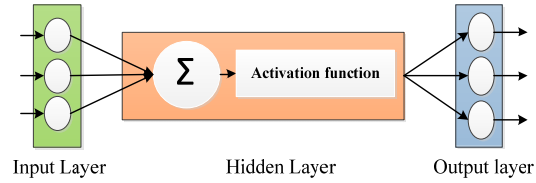
**Figure 3:** A basic of artificial neural network model

The performance of training of ANNs among the processing neurons is based on particular parameters, i.e., weighting coefficients by adjusting their values, till target values matches with the model output. Weights and bias values are modified to minimize the error between model output and target output. Figure 3 defines the basic architecture of ANN, which comprises of three layers: Input layer, hidden layer and output layer. Mapping of information from input layer to output layer is mapped through hidden layer. Each neuron passes its output to the neuron of the higher/next layer and acquires the next input from the lower/previous layer. The selection of number of neurons in the input and output layer is depend upon the nature of problem. The benefit of employing ANNs relates to their capability to appropriately deal massive and complex systems with numerous interdependent parameters [54]. ANNs disregard surplus data with little worth or information and focus on more significant inputs.

### 2.4 Performance measures of Solar Energy Predictors

Performance evaluation is employed by means of root mean square error (RMSE), mean absolute error (MAE), and R2_score with the following formulae:

$$RMSE = \sqrt{\frac{1}{n}\sum_{i=1}^{n}\left(y_{forecasted} - y_{observed}\right)^2} \quad (8)$$

$$MAE = \frac{1}{n}\sum_{i=1}^{n}\left|y_{forecasted} - y_{observed}\right| \quad (9)$$

$$R2\_score = 1 - \frac{\sum\left|y_{forecasted} - y_{observed}\right|}{\sum\left|y_{forecasted} - y_{mean}\right|} \quad (10)$$

The optimal prediction values of performance measures should be zero for MAE, RMSE and one for R2_score.

### 3 Results and Discussion

After data preprocessing, various machine learning models are implemented on it and a series of experiments has been executed on dataset with two grid sizes. Parameter setting of each machine learning model is presented in Table 2, where parameters are selected through trial and error.

### 3.1 Performance Analysis of Proposed ML Models

Results of predicted solar energy from proposed methodologies is compared with actual energy as illustrated in Figure 4 for solar dataset B. This comparison of predicted and actual solar power plots demonstrates a good match. However, statistical error measures of MAE, RMSE and R2_score are numerically provided in Table 3 for further evaluation of each model and relative comparison. Initially, simple base line linear regression model has been evaluated by taking inputs of solar data A and B as explained in section 2.2. Then all the rest of ML techniques are employed.

### 3.2 Comparison of ML Techniques in terms of Error Measures

It can be observed from Table 3 that among variants of each regressor, linear regression-1, ridge regression-II, lasso-III,

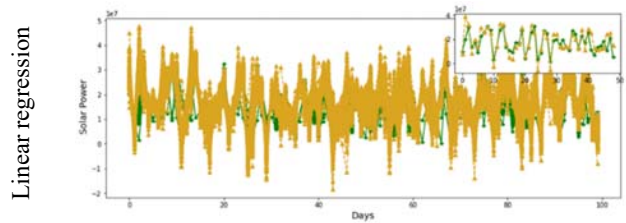

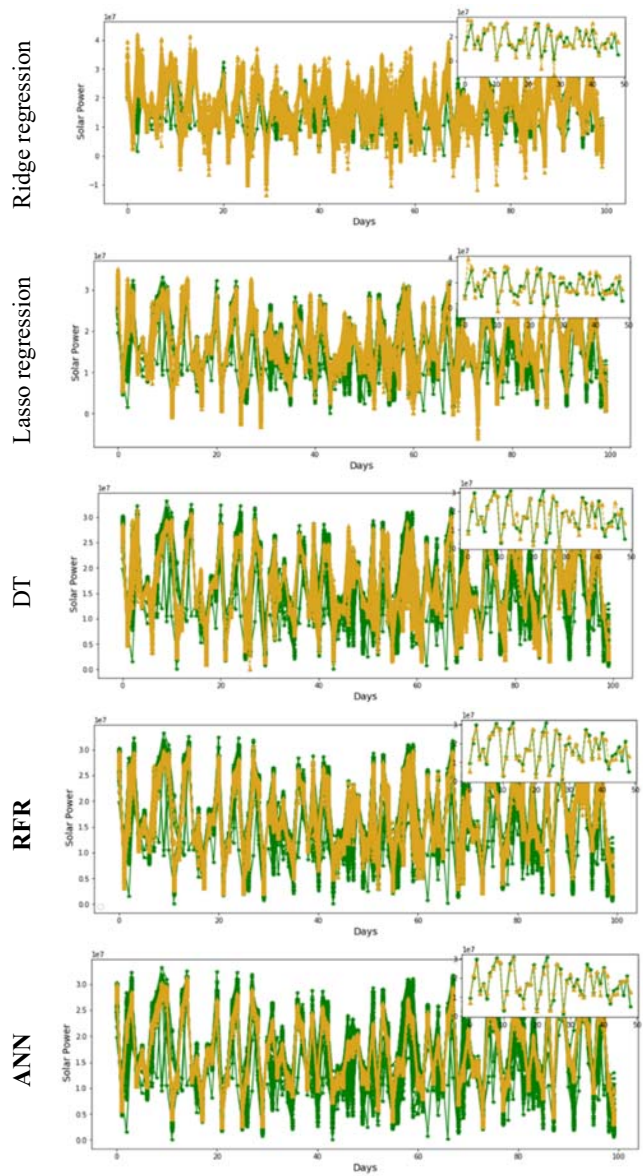

**Figure 4:** Actual and predicted solar energy plots for all solar stations for all base line regressors

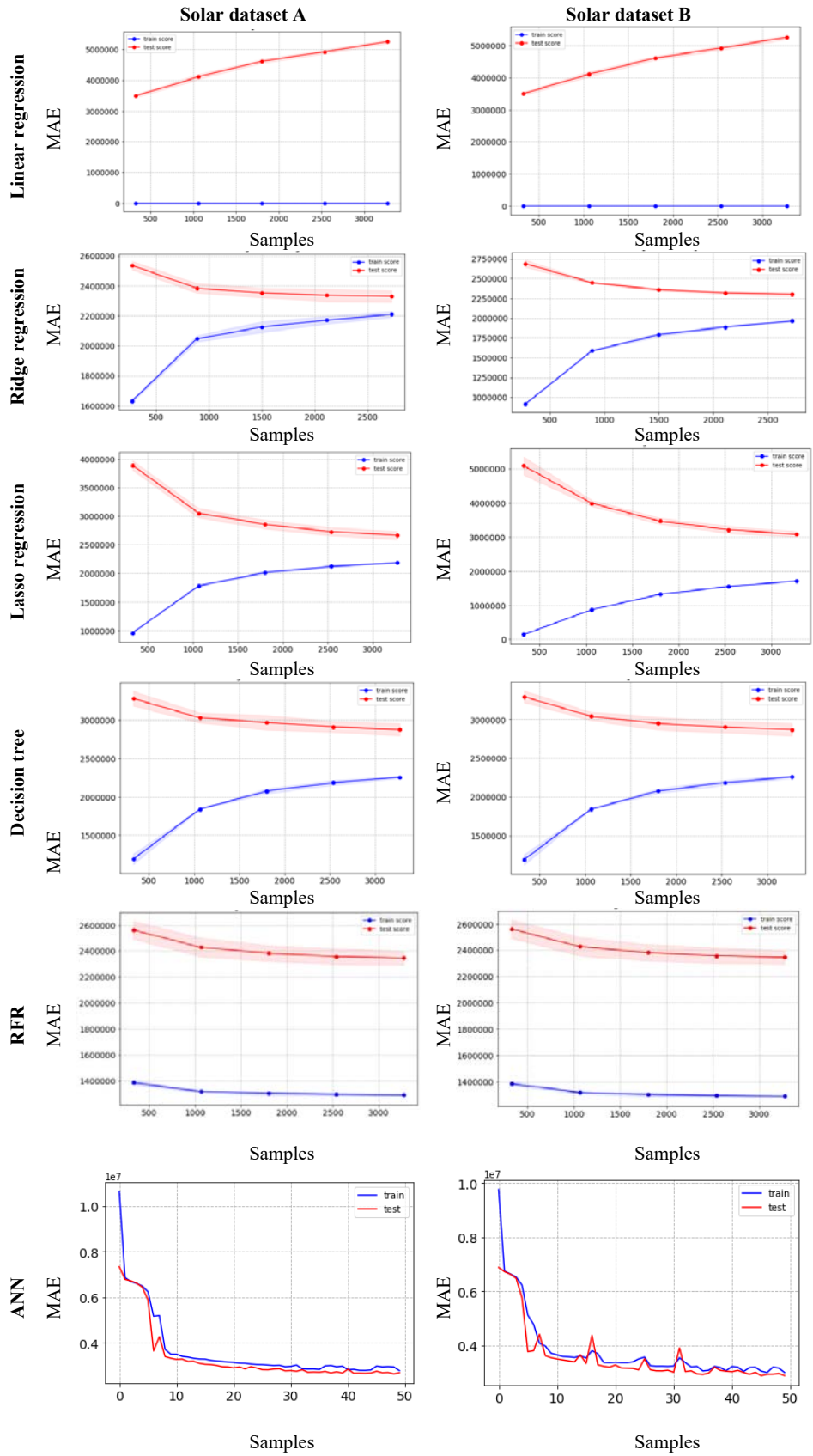

**Figure 5:** Learning curves of solar energy for solar dataset A-B of different techniques

**Table 3:** Performance analysis of solar dataset A and B in terms of parameter optimization

| | Solar dataset A | | | Solar dataset B | | |
|---|---|---|---|---|---|---|
| **Models** | **MAE** | **RMSE** | **R2_score** | **MAE** | **RMSE** | **R2_score** |
| Linear regression-I | **4880032.24** | **6798604.39** | **0.2638796** | **4820050.11** | **6616022.41** | **0.3025184** |
| Linear regression-II | 5486982.49 | 7252964.41 | 0.1621314 | 4820050.12 | 6616022.42 | 0.3025184 |
| Ridge regression-I | 2398962.75 | 3376774.70 | 0.8183589 | 2238475.20 | 3181046.68 | 0.8387703 |
| Ridge regression-II | **2283598.75** | **3220330.28** | **0.8347891** | **2220302.16** | **3142717.47** | **0.8426348** |
| Ridge regression-III | 2232447.59 | 3146332.09 | 0.8422805 | 2234107.73 | 3141788.84 | 0.8427295 |
| Lasso regression-I | 2920022.00 | 4022538.93 | 0.7422443 | 2535824.28 | 3503358.04 | 0.8043779 |
| Lasso regression-II | 2919462.23 | 4021716.91 | 0.7423496 | 2535683.46 | 3503142.84 | 0.8044020 |
| Lasso regression-III | **2918064.42** | **4019665.79** | **0.7426123** | **2535332.05** | **3502605.70** | **0.8044621** |
| Decision tree-I | 3501301.25 | 4600911.51 | 0.6624358 | 3501301.25 | 4600911.51 | 0.6624358 |
| Decision tree-II | **2772250.97** | **3946272.34** | **0.7517212** | **2769557.46** | **3948931.72** | **0.7514041** |
| Random forest-I | **2275017.03** | **3245764.16** | **0.8320232** | **2253614.49** | **3252028.09** | **0.8313877** |
| Random forest-II | 2339102.77 | 3317371.13 | 0.8217654 | 2323285.10 | 3312148.55 | 0.8223654 |
| Random forest-III | 2316722.90 | 3284232.78 | 0.8252453 | 2292778.85 | 3275970.93 | 0.8269548 |
| Artificial neural network-I | **2651832.01** | **3775846.00** | **0.7726286** | **2443559.08** | **3696330.81** | **0.7778733** |
| Artificial neural network-II | 6441782.03 | 7522451.15 | 0.0825467 | 6579833.38 | 7667738.49 | 0.0468257 |

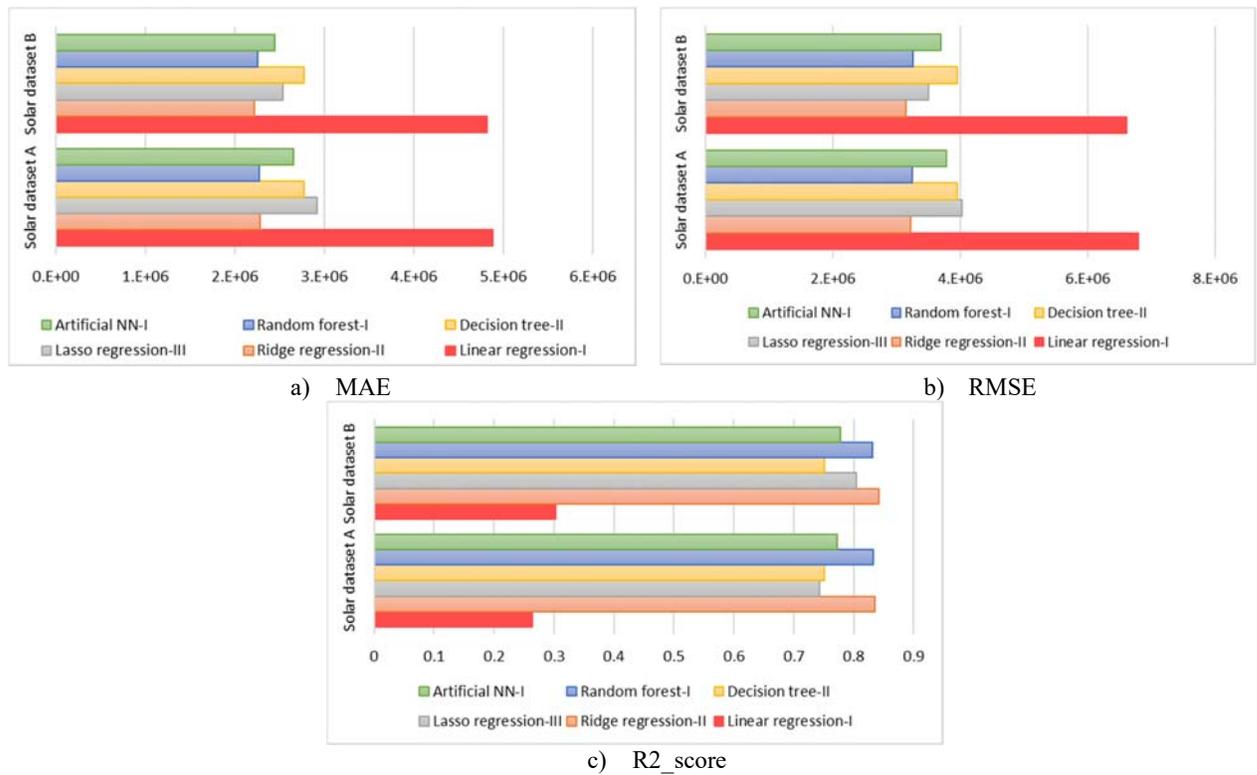

a) MAE   b) RMSE   c) R2_score

**Figure 6:** Error Indices of dataset A and B of the all base line regressors

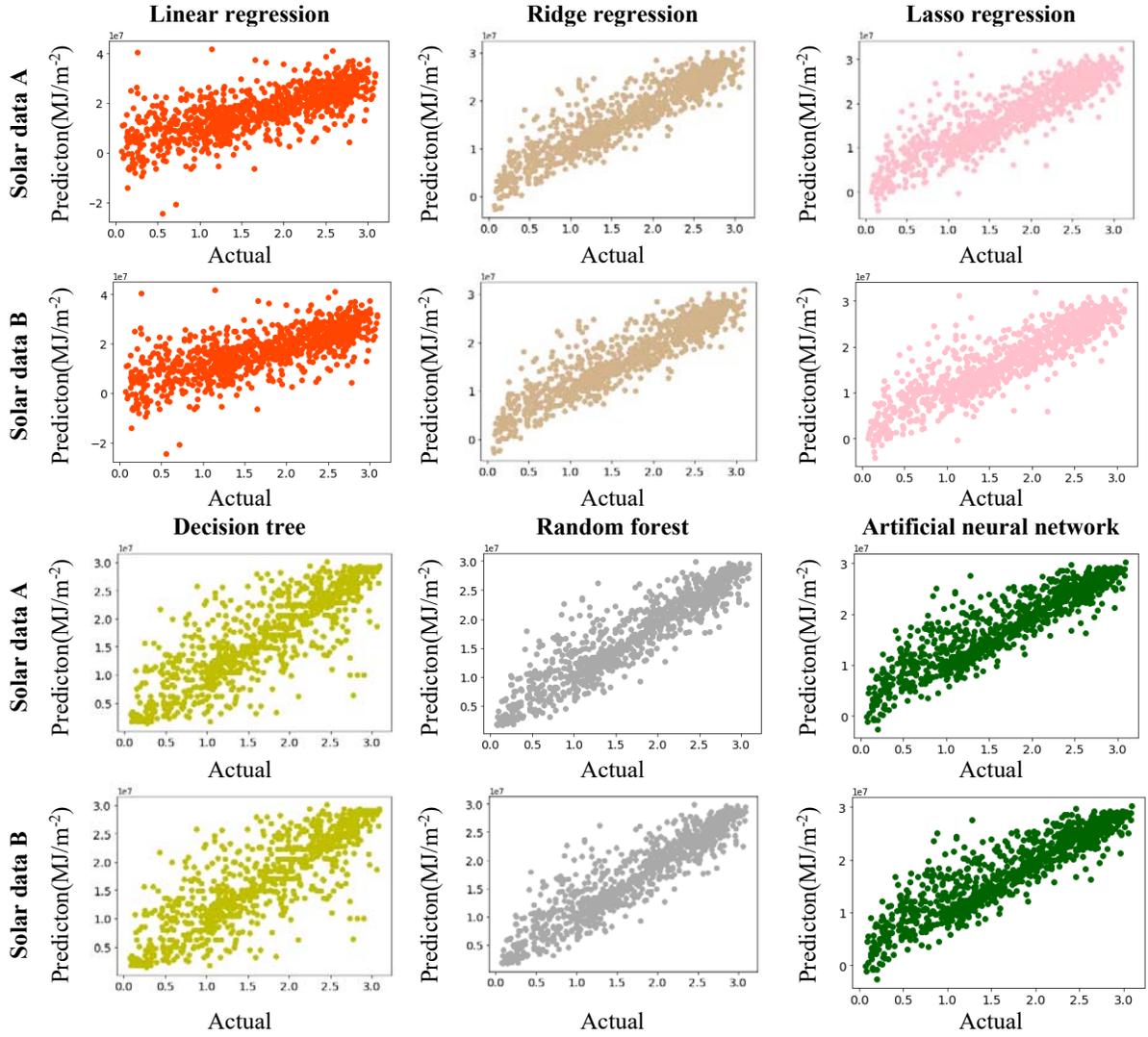

**Figure 7:** Scatter plot of base line regressors of one solar station of solar dataset A-B

**Table 4:** Performance analysis of a single solar station in terms of error measures

| Models | MAE | RMSE | R2_score |
|---|---|---|---|
| Linear regression-I | 4360158.37 | 5958222.87 | 0.432 |
| Linear regression-II | 5031607.95 | 6563079.72 | 0.310 |
| Ridge regression-I | 2339788.09 | 3272217.70 | 0.829 |
| Ridge regression-II | 2222224.64 | 3114850.86 | 0.845 |
| Ridge regression-III | 2190927.10 | 3058312.01 | 0.850 |
| Lasso regression-I | 3151766.16 | 4312270.08 | 0.702 |
| Lasso regression-II | 3149185.54 | 4308681.62 | 0.703 |
| Lasso regression-III | 3142768.75 | 4299745.58 | 0.704 |
| Decision tree-I | 3269495.07 | 4351369.72 | 0.697 |
| Decision tree-II | 2633471.37 | 3959846.00 | 0.749 |
| Random forest-I | 2314902.76 | 3294124.19 | 0.826 |
| Random forest-II | 2200779.76 | 3148685.63 | 0.841 |
| Random forest-III | 2199231.52 | 3145867.71 | 0.842 |
| Artificial neural network-I | 2752639.79 | 3759635.83 | 0.774 |
| Artificial neural network-II | 6379583.48 | 7505634.24 | 0.098 |

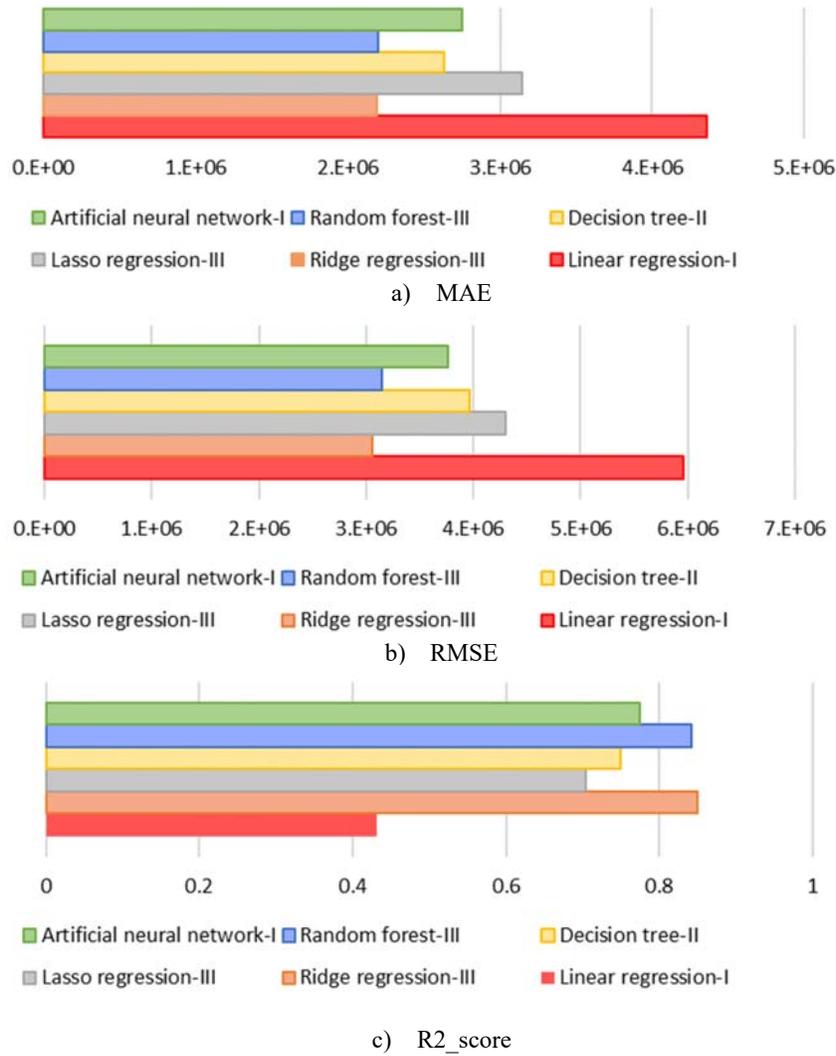

a) MAE

b) RMSE

c) R2_score

**Figure 8:** Performance of a single solar station as bar plots

**Table 5:** Performance analysis of regression models with Aggarwal *et* al. [35] in terms of MAE

| Model | Solar dataset |
|---|---|
| Random normal | 9201211.85 |
| Gaussian mixture | 4019469.94 |
| Spline Interpolation | 2611293.30 |
| LSR | 2524141.80 |
| Regularized LSR | 2544190.39 |
| FFNN | 2520991.85 |
| Ensemble-FFNN only | 2416987.41 |
| Ensemble-FFNN and LSR | 2411989.61 |
| Linear regression-I | 4880032.24 |
| Ridge regression-II | 2283598.75 |
| Lasso regression-III | 2918064.42 |
| Decision tree-II | 2772250.97 |
| Random forest-I | 2275017.03 |
| Artificial neural network-I | 2651832.01 |

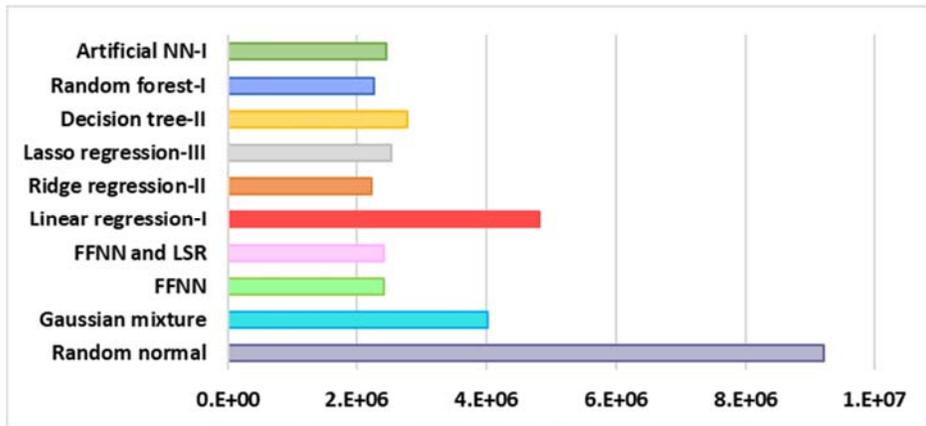

a) Comparison of MAE from various models

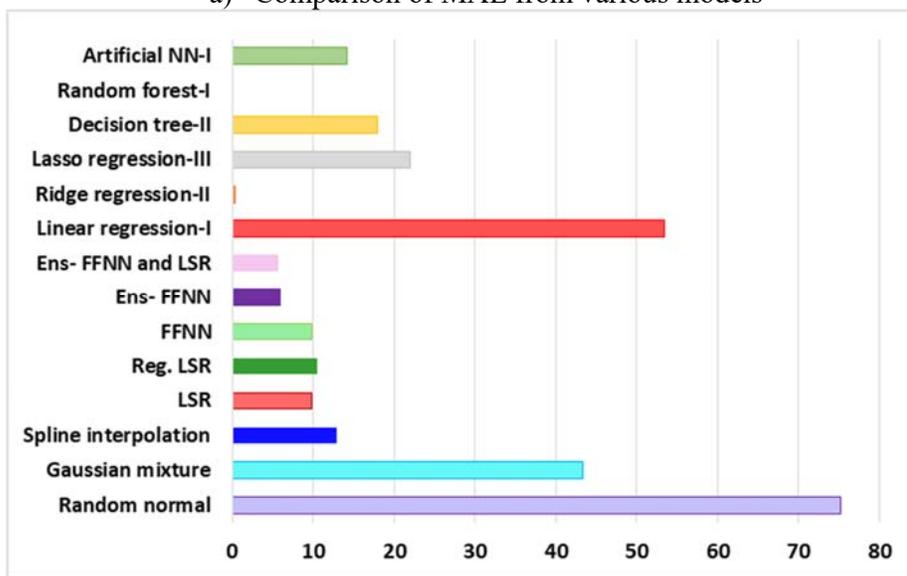

b) Percentage improvement of MAE from RFR with existing techniques

**Figure 9:** Comparison of various models in terms of (a) MAE and (b) Percentage MAE

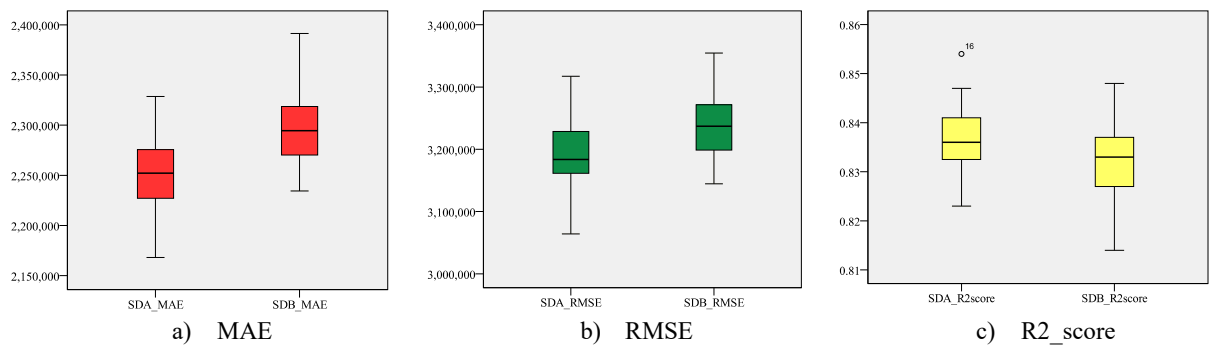

a) MAE        b) RMSE        c) R2_score

**Figure 10:** Boxplots for performance measures of Ridge regression-II on independent multiple runs

decision tree-II, random forest-I and ANN-I have demonstrated the best error measures in terms of MAE, RMSE and R2_score for both solar datasets A and B. As far as MAE is concerned, it can be noticed that among these regression models, random forest- I performs best with MAE 2275017.03 and ridge regression-II with MAE of 2283598.75 on solar dataset A. Moreover, on solar dataset B, ridge regression-II with MAE of 2220302.16 is better as compared to random forest-I with MAE of 2253614.49. However, it has been observed that due to various characteristics of solar stations, a vast variety of changes over statistical error measures at different stations is seen, as demonstrated in Table 3.

The performance of the proposed machine learning models is also carried out with respect to learning curves on solar dataset A and B both for training and testing, subfigures are given in Figure 5. These error indices are displayed as bar charts for better observation in all subfigures of Figure 6. It can be noticed the all performance indices more or less behave in the same manner for both datasets.

### 3.3 Actual versus Predicted Daily Solar Power

Scatter plots of actual versus predicted daily solar power have been plotted for both datasets for all proposed machine learning techniques. It can be realized that predicted power is in a good pattern with the actual wind power, which is demonstrated in Figure 7. Overall trend can be easily observed from these scatter plots.

### 3.4 Performance Analysis on a Single Solar Station

To check the model generalization ability, performance analysis is measured on single solar station with respect to all proposed methods and is illustrated in Table 4. While Figure 8 shows the best performed methodologies in form of bar charts. It almost illustrates similar behavior as of a grid of many solar stations.

### 3.5 Comparison with Existing Techniques

Additionally, accuracy and precision is evaluated on the basis of comparison with existing techniques [35]. The values of MAE and percentage improvement of MAE from RFR are compared with random normal, Gaussian mixture, FFNN and ensemble model of FFNN and LSR of existing techniques and other proposed models as illustrated in Table 5 and Figure 9 for test data of solar energy. It can be observed that some improvements with ridge regression and RFR are due to obvious advantage of non-linear predictors. Percentage improvement of MAE from RFR ranges from 5.67% upto 75 % of existing techniques.

### 3.6 Performance Analysis on Multiple Independent Runs

In order to evaluate whether the proposed models are stable, reliable, and convergent on the basis of performance indices, ridge regression-II for solar dataset A and B have been run independently 60 times. The boxplot results of MAE, RMSE and R2_score are presented in Figure 10 for multiple independent runs for both datasets. It can be observed that boxes do not overlap one another; and therefore there is a difference between two groups.

## 4 Conclusion

AMS 2013-14 solar energy prediction dataset has been taken, where non-linear regression mosels, Linear regression, Ridge regression, Lasso regression, DT, RFR and ANN. It can be concluded that random forest and then ridge regression model outperformed all techniques for dataset A and B. ridge and random forest give MAE values of 2283598.75 and 2275017.03, respectively. It can be examined that choosing smaller grid size (solar dataset B) enhanced the performance. It can also be concluded from error indices that the same techniques demonstrate enhanced performance on a single solar station as well. In order to analyze the robustness and efficacy of proposed models comparison with existing techniques on dataset A has been carried out, which demonstrates better performance of random forest and ridge regression. Multiple independent runs demonstrate the stability and convergence of the proposed scheme.

**Acknowledgements**

Authors acknowledge PR Lab at department of Computer & Information Sciences, Pakistan Institute of Engineering and Applied Sciences for the computational resources.

**Declaration of interests**

☒ The authors declare that they have no known competing financial interests or personal relationships that could have appeared to influence the work reported in this paper.